\documentstyle[aps,12pt]{revtex}
\begin{document}

\title{First-principles Calculation of the Formation Energy
in MgO--CaO Solid Solutions}

\author{S. V. Stolbov$^{(1)}$, R. E. Cohen$^{(1,2)}$}
\address{$^{(1)}$Carnegie Institution of Washington, Washington,
DC 20015}
\address{$^{(2)}$Seismological Laboratory, California Institute
of Technology, Pasadena, CA 91125}

\maketitle

\begin{abstract}
The electronic structure and total energy were calculated for ordered
and disordered MgO-CaO solid solutions within the multiple scattering
theory in real space and the local density approximation. Based on the
dependence of the total energy on the unit cell volume the equilibrium
lattice parameter and formation energy were determined for different
solution compositions. The formation energy of the solid solutions is
found to be positive that is in agreement with the experimental phase
diagram, which shows a miscibility gap. 
\end{abstract}

%%\twocolumn

\section{Introduction}
In the last decade significant progress was achieved in
application of first principles methods to calculate properties
of materials. The full-potential linearized-augmented-plane-wave
(FLAPW) and ultra-soft pseudopotential methods allow one to map
phonon potential surface, describe phase transitions,
electromechanical properties of ferroelectrics
\cite{REC92,Fu,KSV} and magnetic properties of materials
\cite{Singh}. However these methods are applicable to
ordered systems with translational symmetry, whereas many
technologically relevant materials do not meet this requirement.
Such important properties as ultrahigh piezoelectric efficiency,
colossal magnetoresistance, high oxygen permeation through fuel
cell membranes occur only when corresponding oxides form solid
solutions, most of which are disordered.

First-principles methods developed for disordered systems are
mostly based on the Green's function formalism such as the
Korringa-Kohn-Rostoker (KKR) method \cite{korr,krost} that allows 
one to take into account atomic disorder within the average 
$t$-matrix (ATA) \cite{e-s,Bansil} or coherent potential (CPA) 
\cite{soven,taylor} approximations. They have been successfully 
applied to metallic alloys \cite{johnson86,johnson90,pinski91,jordan,Abr}. 
A few calculations were performed for copper oxides \cite{Sz,SS} 
with mixed covalence -- ionic character of chemical bonding. In the
present paper, we consider the simplest oxide solid solution, the
system MgO-CaO. Magnesiow\"{u}stite (Mg,Fe)O is believed to be a
major constituent of the Earth's lower mantle, so that it would be
an ideal system for study, but the FeO constituent leads to complex
Mott insulator behavior. The study of CaO in MgO can be considered
as a step in the direction of the understanding of solid solution in
minerals, as well as important in understanding the behavior of the
minor element Ca. Understanding of oxide solid solutions is also
important from the perspective of ferroelectric solid solutions and
high temperature superconductors.

Several approaches have been used previously to study the MgO-CaO
system. The electronic structure and total energy were calculated
for constitutive compounds of the system, MgO and CaO using FLAPW
method \cite{MRECK}. A potential model \cite{JACS} was developed
to compute the phase diagram of the MgO-CaO solid solution. However
the approach was found to be inadequate to reproduce the properties
with required accuracy. Authors of Ref. \onlinecite{KC} applied a
tight-binding method to calculate the formation energy of ordered
MgO-CaO solutions. Tight-binding parameters were obtained from
results of first principles pseudopotential calculations performed
for MgO, CaO and MgCaO$_2$ compounds and then used for computing
the total energy of solutions with different compositions and crystal
structures. Comparison with the pseudopotential results suggests that
the tight-binding approach is an efficient tool to study ionic
oxides. However those were still ordered systems.

In the present paper, we considered both ordered and fully disordered
MgO-CaO solutions using the multiple scattering theory within the local
density approximation. We calculated the electron structure, total
energy, equilibrium lattice parameters and formation energy for
ordered MgO, CaO and MgCaO$_2$ compounds as well as for 4 disordered
MgO-CaO solutions with different compositions. The obtained formation
energy values were compared to results of Ref. \onlinecite{KC}.

\section{Computational Method}
The calculations were performed by means a computer code based on the
local density approximation (LDA) \cite{KSh} and multiple scattering
theory. The code embodies the local self-consistent multiple
scattering (LSMS) method \cite{Stocks95} where a compound is divided
into overlapping clusters -- local interaction zones (LIZs) centered
around atoms of different sorts. The multiple scattering problem is
solved for each LIZ separately in the {\it lattice site -- angular
momentum} representation for the {\it muffin-tin} (MT) potential
given the cluster Green's functions, the local densities of
electron states and valence charge density for central atom.

Our approach is self-consistent with respect to local charge
densities and potentials. However it does not treat the disorder
self-consistency. We use an actual single site scattering matrix
$t^i_l(E)$ for the central atom of LIZ and average $t$-matrix:
$\tilde{t^i_l}=xt^a_l+(1-x)t^b_l$ for surrounding sites, if they
belong to a sublattice with a substitutional disorder of the $a$-
and $b$-sort atoms with concentrations $x$ and $1-x$. According
to the Ref. \onlinecite{Bansil}, such an approach is more accurate
than the regular average $t$-matrix approximation. In the
considered solid solution the Mg and Ca atoms randomly occupy
the metal sublattice sites. They have the same number of
valence electrons. This also allows us to believe that our
approximation is reasonable for the system. One more advantage
of the approach is it can be explicitly implemented within the
ordinary LSMS theory \cite{Stocks95}.

Solving the Poisson equation at each iteration, we use the actual
charge density for the local contribution to the MT potential
and the average charges for the contribution of the substitutionally
disordered sublattice to the Madelung potential. The exchange and
correlation parts of the potential is determined within LDA using
the technique described in Ref. \onlinecite{GunLun}.

The total energy was calculated following procedure described in Ref.
\onlinecite{johnson90} and using an expression for multicomponent
compound \cite{Schm}.

In the present calculations, each LIZ built around Mg, Ca and O
sites contained 123 atoms. This provided good convergence over
the LIZ size, since the difference in the total energies
obtained for 123 and 93 atom LIZs was 3 and 5 mRy for MgO and
CaO respectively. To estimate an error induced by MT potentials,
the calculations were performed in two approximations --
maximum MT radii proportional to the unit cell volume and fixed
MT radii corresponding to a minimum volume considered for a
given compound.

\section{Results and Discussion}
The density of electron states, charge density and total energy were
calculated versus the unit cell volume for ordered compounds MgO, CaO
and MgCaO$_2$ and disordered solid solutions Mg$_{1-x}$Ca$_x$O with
$x=0.2, 0.5, 0.7, 0.8$. For MgCaO$_2$ the structure was chosen where
cations were ordered by an alternate stacking of Mg and Ca planes
along the [001] direction. Thus the cations form the Ll$_0$ structure
in Strukturbericht notation \cite{Stbr}.

Based on the computed volume dependence of the total energy the
equilibrium volumes were calculated for each considered compound.
The results are shown in Fig. 1. The concentration dependence of unit
cell volume is found to be close to linear one, though some wavy
behavior is seen. The difference in results obtained by means of
maximum and fixed MT radii approaches gives an order of magnitude for
the error induced by {\it muffin-tin} approximation. For the MgO and
CaO compounds the calculated equilibrium volumes are less than
experimental ones that reflects the well known feature of LDA.

Using the calculated values of total energy we have determined the
formation energy of the solutions determined as:
$$ E_f = E(Mg_{1-x}Ca_xO) -[(1-x)E(MgO)+xE(CaO)],$$
where E is the total energy of corresponding compound. The $E_f$
values compared to results of pseudopotential calculations
\cite{KC} are shown in Fig. 2. Our result and result of Ref.
\onlinecite{KC} obtained for the ordered Ll$_0$ phase of MgCaO$_2$
are in a good quantitative agreement (0.285 eV and 0.282 eV
respectively). The difference in formation energy of the ordered and
disordered MgCaO$_2$ is less than 10\%. Such a small difference
takes place since a decay of band states caused by random potentials
mostly involves unoccupied states in these ionic compounds  and is
much lower than in case of materials with covalence bonding \cite{SS}.
The formation energy is found to be positive for all considered
solid solutions suggesting that a phase separation is preferable
for this system. This is in agreement with the experimental phase
diagram \cite{YA}, which shows a miscibility gap.

Charge transfer is one of the key mechanisms determining the total
energy of complex oxides. Therefore, we have calculated effective
charges on atoms in the considered solid solutions by integration
of the valence electron density over Wigner-Seitz spheres. It is
important at this stage to make a clear reasonable definition of
the space belonging each nonequivalent atom that is determined
by the ratio of the Wigner-Seitz radii $r^{WS}_O/r^{WS}_M$
(subscripts $O$ and $M$ denote oxygen and metal respectively).
We suppose that in the present case it is convenient for
interpretation purpose to keep this ratio independent of the
metal composition. This ratio should also make some physical
sense. Because the considered solutions are definitely ionic,
the ratio should be related to the ionic radii. Thus we come
to the Wigner-Seitz radii ratio defined as
$$ r^{WS}_O/r^{WS}_M=2R^{OII}_i/[R^{MgII}_i+R^{CaII}_i]=1.24,$$
where $R^{OII}_i,R^{MgII}_i$ and $R^{CaII}_i$ are ionic radii
of O, Mg and Ca respectively. We have also determined the average
cation charges in the solutions as $Q_{av}=(1-x)Q_{Mg}+xQ_{Ca}$.
The results are shown in Fig. 3. Since the Ca atom has the
valence wave functions more extended than Mg, substitution of
Mg with Ca leads to a noticeable increase in the electron charge
on oxygen atoms (more than 0.2e, going from MgO to CaO). The
concentration dependence of these charges as well as average
cation charges are found to be linear.

In summary, we have calculated the electron structure, total energy,
equilibrium lattice constants and formation energy for ordered and
disordered solid solutions Mg$_{1-x}$Ca$_x$O. A linear composition
dependence has been found for lattice parameters and effective
charges. The formation energy is positive for all considered
materials that is in agreement with the experimental phase diagram.
The results suggest that LSMS method and LDA can be an efficient
tool to study properties of disordered ionic solutions.

{\bf Acknowledgements.}
This work was supported by Office of Naval Research grant
N00014-97-1-0052. Computations were performed on the Cray
SV1 at the Geophysical Laboratory, supported by NSF grant
EAR-9975753 and the W.\ M.\ Keck Foundation.

\begin{figure}
\caption
{The composition dependence of the equilibrium
volume per oxygen atom calculated with maximum MT radii
(squares) and fixed MT radii (circles). The dot-dashed
line connects the experimental volume values obtained
for stoichiometric MgO and CaO compounds.}
\end{figure}

\begin{figure}
\caption
{The formation energies calculated in the present
work for disordered Mg$_{1-x}$Ca$_x$O solid solutions
(triangles), ordered MgCaO$_2$ compound (circle), as well
as obtained by means of pseudopotential method \protect\cite{KC}
for the following ordered phases: $x=0.25$ (CaMg$_3$O$_4$),
higher cross -- $Ll_2$ structure, lower cross -- $DO_{22}$
structure; $x=0.33$ (CaMg$_2$O$_3$) -- MoPt$_2$ structure;
$x=0.5$ (CaMgO$_2$), from higher to lower cross -- $Ll_0$,
$Ll_1$ and $A_2B_2$ structures respectively; $x=0.67$
(Ca$_2$MgO$_3$) -- MoPt$_2$ structure; $x=0.75$
(Ca$_3$MgO$_4$) higher cross -- $Ll_2$ structure, lower cross
-- $DO_{22}$ structure; $x=0.8$ (Ca$_4$MgO$_5$) -- Ni$_4$Mo
structure.}
\end{figure}

\begin{figure}
\caption
{The composition dependence of Wigner-Seitz charges on
Mg (up triangles), Ca (down triangles), O (circles) and average
cation charges (squares) in Mg$_{1-x}$Ca$_x$O.}
\end{figure}

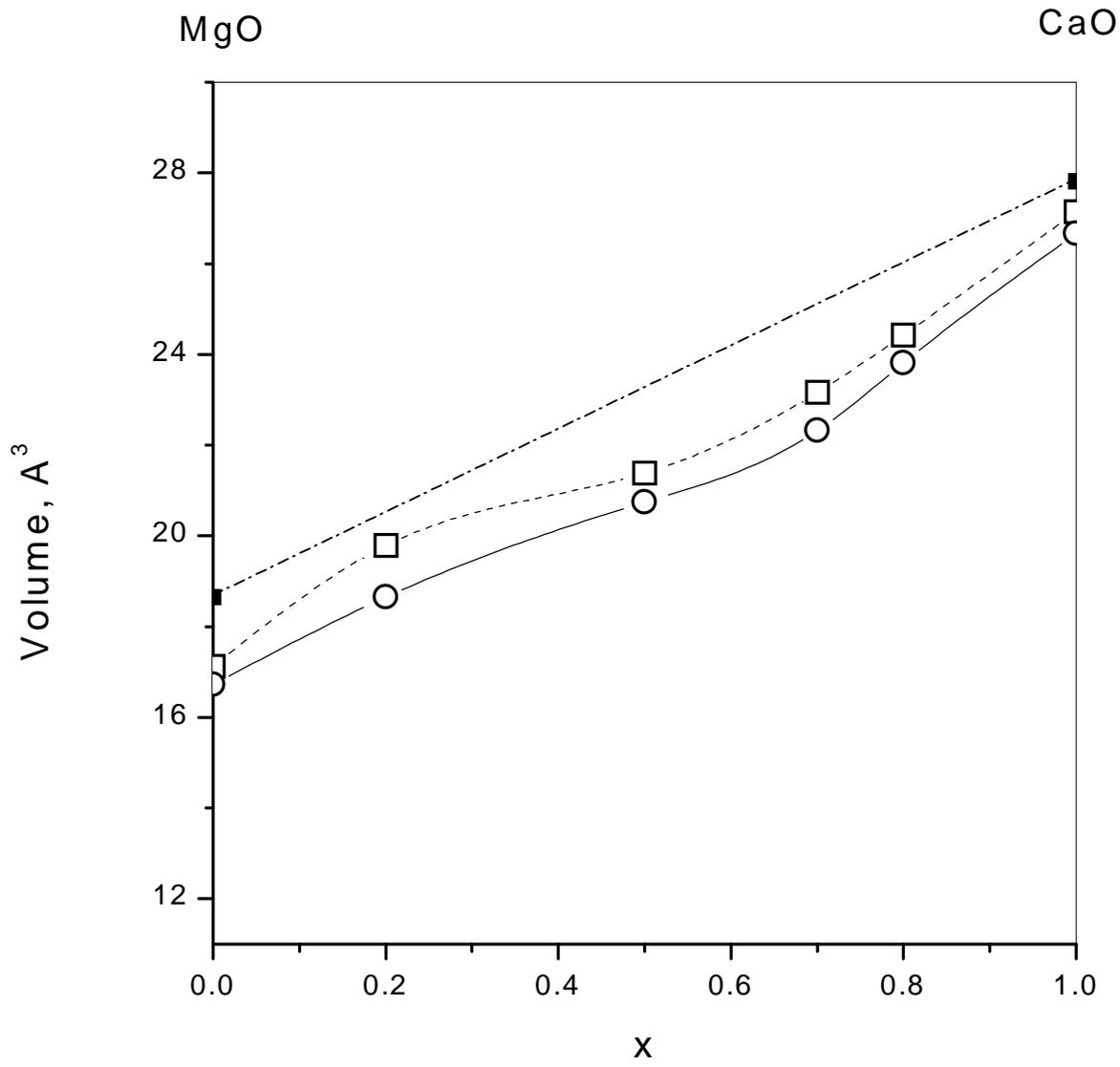
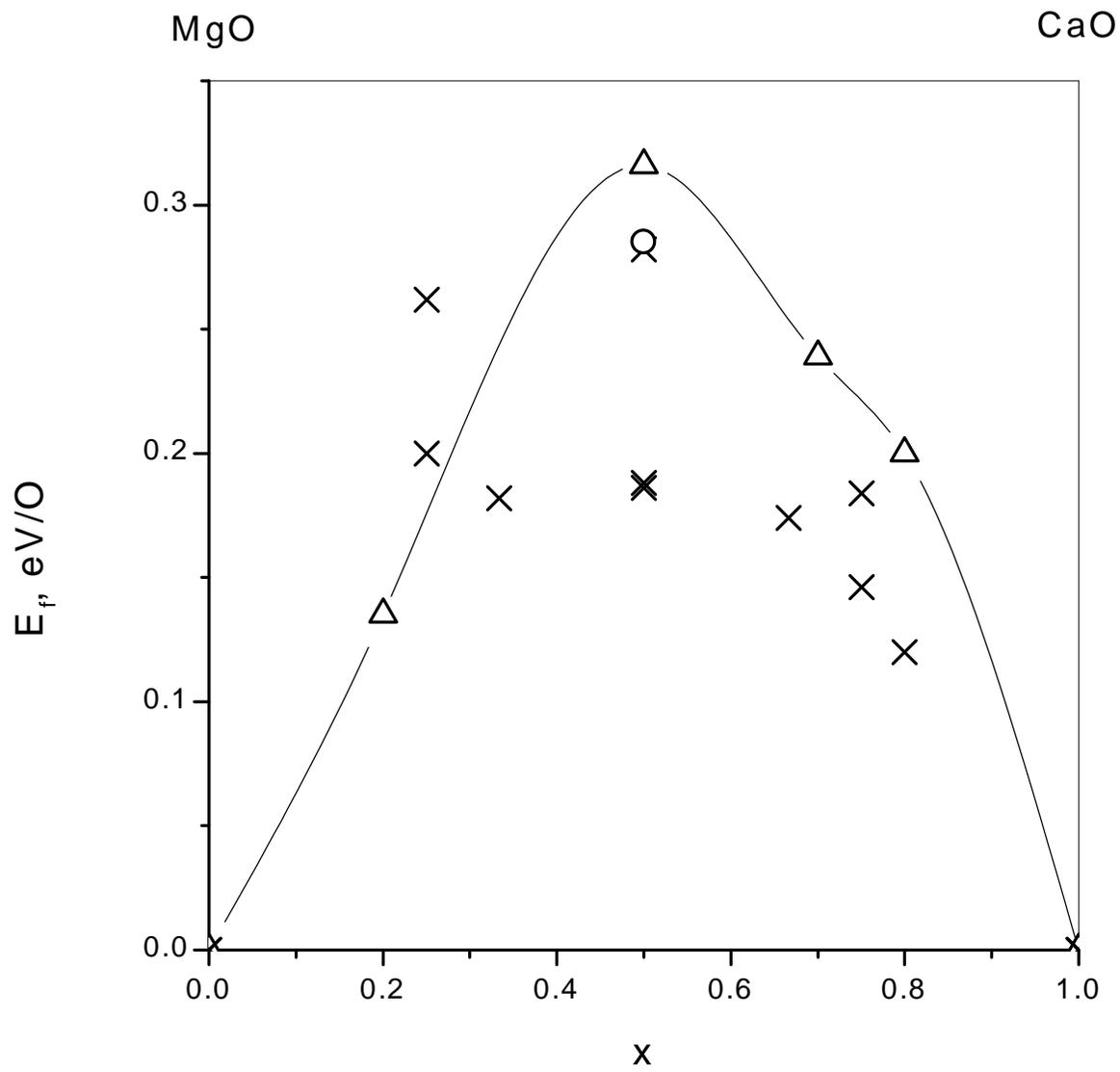
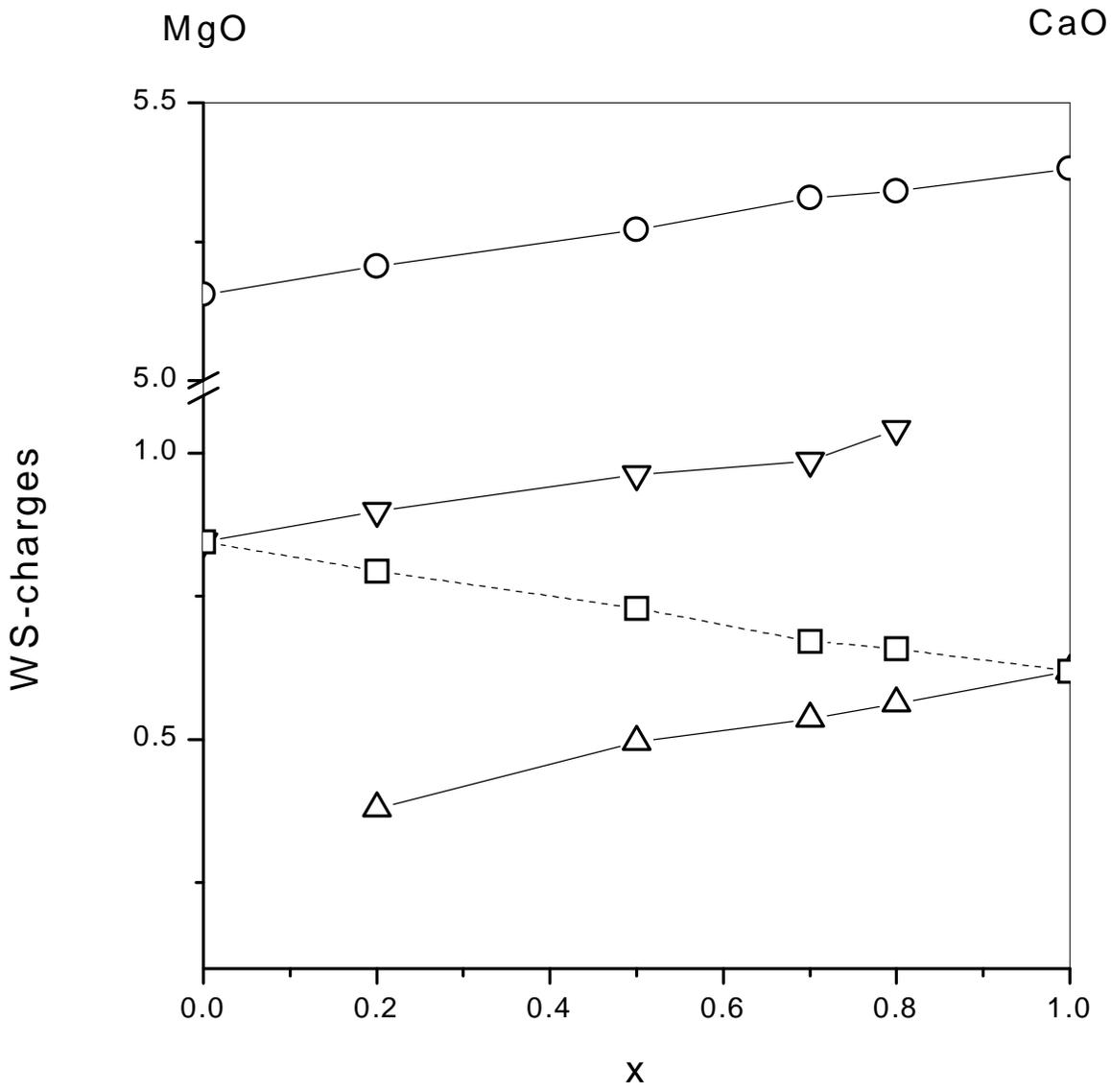
\end{document}